%% file: ms.tex
\documentclass{sig-alternate-10pt}
\usepackage{wrapfig,graphicx,lipsum}
\usepackage{epsfig,endnotes}
\usepackage{amssymb}
\usepackage{amsmath}
\usepackage{amsfonts}
\usepackage{cite}
\usepackage{booktabs}
\usepackage{colortbl}
\usepackage{indentfirst}
\usepackage{subfigure}
\usepackage{tabularx}
\usepackage{graphicx}
\usepackage{color}
\usepackage{xspace}
\usepackage{thumbpdf}
\usepackage{hyperref}
\usepackage{acmcopyright}
\usepackage{listings}
\usepackage{verbatim}
\usepackage{colortbl}
\usepackage{caption}
\usepackage{lipsum}
\usepackage{fancyhdr} 

\usepackage{url}

\newcommand\sysname{Dyloc}
\newcommand{\argmin}{\mathop{\mathrm{argmin}}}
  
\author{
Ouyang Zhang and Kannan Srinivasan\\
  Department of Computer Science and Engineering  \\
      The Ohio State University, Columbus, OH 43210}
\begin{document}
%\CopyrightYear{2016} 
%\setcopyright{acmcopyright}
%\conferenceinfo{CoNEXT '16,}{December 12-15, 2016, Irvine, CA, USA}
%\isbn{978-1-4503-4292-6/16/12}\acmPrice{\$15.00}
%\doi{http://dx.doi.org/10.1145/2999572.2999582}

\clubpenalty=10000 
\widowpenalty = 10000

%\date{}
\graphicspath{{Figures/}}

\title{\vspace*{-0cm} \sysname:\ Dynamic and Collaborative User-controlled AOA based Localizing System with your laptops
\vspace*{-0.6cm}}

%\title{\LARGE \bf \sysname:\ User-friendly Fine-grained Gesture Recognition  using WiFi Signals}
%\title{\vspace*{-5cm}  \LARGE \bf \sysname:\ User-friendly Fine-grained Gesture Recognition  using WiFi signals\vspace{-0.9cm}}
%\author{
%\normalfont{\Large Paper \#: 41}}

%\author{
%Ouyang Zhang and Kannan Srinivasan\\
%  Department of Computer Science and Engineering  \\
%      The Ohio State University, Columbus, OH 43210\\
%	zhang.4746@buckeyemail.osu.edu, kannan@cse.ohio-state.edu}
%\author{
%Bo Chen$^{\dagger}$, Vivek Yenamandra$^{\dagger}$ and Kannan Srinivasan\\
%  Department of Computer Science and Engineering  \\
%      The Ohio State University, Columbus, OH 43210\\
%	\{chebo, yenamand, kannan\}@cse.ohio-state.edu\\
%$\dagger$Co-primary Authors}

%%Tarun Bansal$^{\dagger}$, Bo Chen$^{\dagger}$, Kannan Srinivasan and Prasun Sinha\\
%%  Department of Computer Science and Engineering  \\
 %%     The Ohio State University, Columbus, OH 43210\\
%%	\{bansal, chebo, kannan, prasun\}@cse.ohio-state.edu\\
%%$\dagger$Co-primary Authors}

\maketitle
\pagestyle{plain}
%\setcounter{page}{1}
%\pagenumbering{arabic}
% Use the following at camera-ready time to suppress page numbers.
% Comment it out when you first submit the paper for review.
%%\thispagestyle{empty}
%%\graphicspath{{Figures/}}

\input{sec_Abstract}

%%%%%%%%%%%%%%%%%%%%%%%%%%%%%%%%%%%%%%%%%%%%%%%%%%%%%%%%%%
\input{sec_Intro}
\input{sec_Related}
\input{sec_Overview}
% %

\input{sec_SysDesign}

{\footnotesize\bibliographystyle{acm}
\bibliography{citations}}

\end{document}

%% file: sec_Abstract.tex
\vspace*{-2cm}
\subsection*{Abstract}
Currently, accurate localization system based on commodity WiFi devices is not broadly available yet. In the literature, the solutions are based on either network infrastructure like WiFi router, which have at least three antennas, or sacrifice accuracy with coarse-grained information like RSSI. In this work, we design a new localizing system - \sysname\ which is accurate based on AOA estimation and instantly deployable on users' devices. 

\sysname\ is designed to be dynamically constructed with user's devices as network nodes without any network infrastructure. On the platform of laptops, our system achieve comparable localization accuracy with state-of-the-art work despite of the limitation of less number and large separation of antennas. We design multi-stage signal processing to resolve the ambiguity issue arisen in this scenario. To enable dynamic and collaborative construction, our system can accurately conduct self-localization and also eliminate the need of infrastructure anchors, which is due to the dedicated two-layer algorithm design. 

%First, carrier-based sensor enhancement is used to combat the limited number of antennas on laptops. Second, unique observation offers us a narrow beam to precisely locate the target. Last but not least, corporation and synthesized processing on data from multiple laptops help us resolve the ambiguity in candidate location. With \sysname\, the general public could generate a localization environment with their own control which can helps in a big brunch of activity organization. From extensive experiments, \sysname\ can give us a localization accuracy of about 0.5 m within an typical office environment and conference building. 

%
%\subsection*{Keywords}
%Gesture Recognition; WiFi Signals; Signal Cancellation

%% file: sec_Intro.tex
%\vfill\eject 
\section{Introduction}
\label{sec:intro}
%\vspace*{-0.3cm}

In recent decades, RF-based positioning attracts lots of interests and  efforts over the world in both research and industry community. Those works show promising and innovative applications in mobile computing, human-computer interaction and assistant living \cite{mitev2016indoor,Forbes, Frost, MTReview}. In the literature, various hardware platforms have been exploited to design and implement the proposed systems upon, such as software defined radio (SDR) \cite{xiong2014synchronicity, xiong2015tonetrack,joshi2013pinpoint,xiong2013arraytrack}, RFID\cite{ma2017minding,wang2013dude,wang2013rf,wang2014rf}, and commodity WiFi devices\cite{gjengset2014phaser,sen2013avoiding,liu2012push,kotaru2015spotfi,choi2017deep,chintalapudi2010indoor}. Among these, commodity WiFi based system shows great advantage as it can be instantly deployed on existing WiFi devices, providing an ubiquitous service like GPS.

State-of-the-art localizing systems with commodity off-the-shelf WiFi adapters (CoTS WiFi) utilize several measurements reported by the chipset to derive relevant information, such as  RSSI, CSI or AOA information. Fingerprint based \cite{youssef2005horus,sen2012you,nandakumar2012centaur,azizyan2009surroundsense} and RSSI based systems \cite{bahl2000radar,chintalapudi2010indoor,ferris2007wifi} suffer from insufficient modeling of RSSI or  expensive recurring operations on changing environments and hence are either inaccurate (median error of 2-4 m) or difficult to deploy.  By contrast, AOA (angle of arrival) based systems \cite{xiong2013arraytrack,kumar2014accurate,kumar2014lte,sen2013avoiding} can achieve localizing accuracy of tens of centimeters. 

However, existing AOA based systems built on CoTS WiFi utilized WiFi infrastructure like routers only accessible to the network administrator, requiring at least three antennas for the antenna array. One reason is that the resolution of AOA estimation on the antenna array is greatly affected by the number of antennas in the array. Thus, nearly all WiFi devices possessed by normal users and WiFi cards are hard to deploy these systems due to prevalent two-antenna configuration. For another reason, the antenna spacing on commodity device is fixed and hence cannot be adjusted to meet their design specification. For example, \cite{joshi2013pinpoint,wang2013dude,xiong2013arraytrack} requires the antenna spacing to be less than $\lambda/2$. In this work, the goal is to solve the above problem and enable an user-controlled localizing system based on AOA with high accuracy. 
\begin{figure}[t!]
\centering
\vspace*{-1.0cm}
\includegraphics[width=0.4\textwidth]{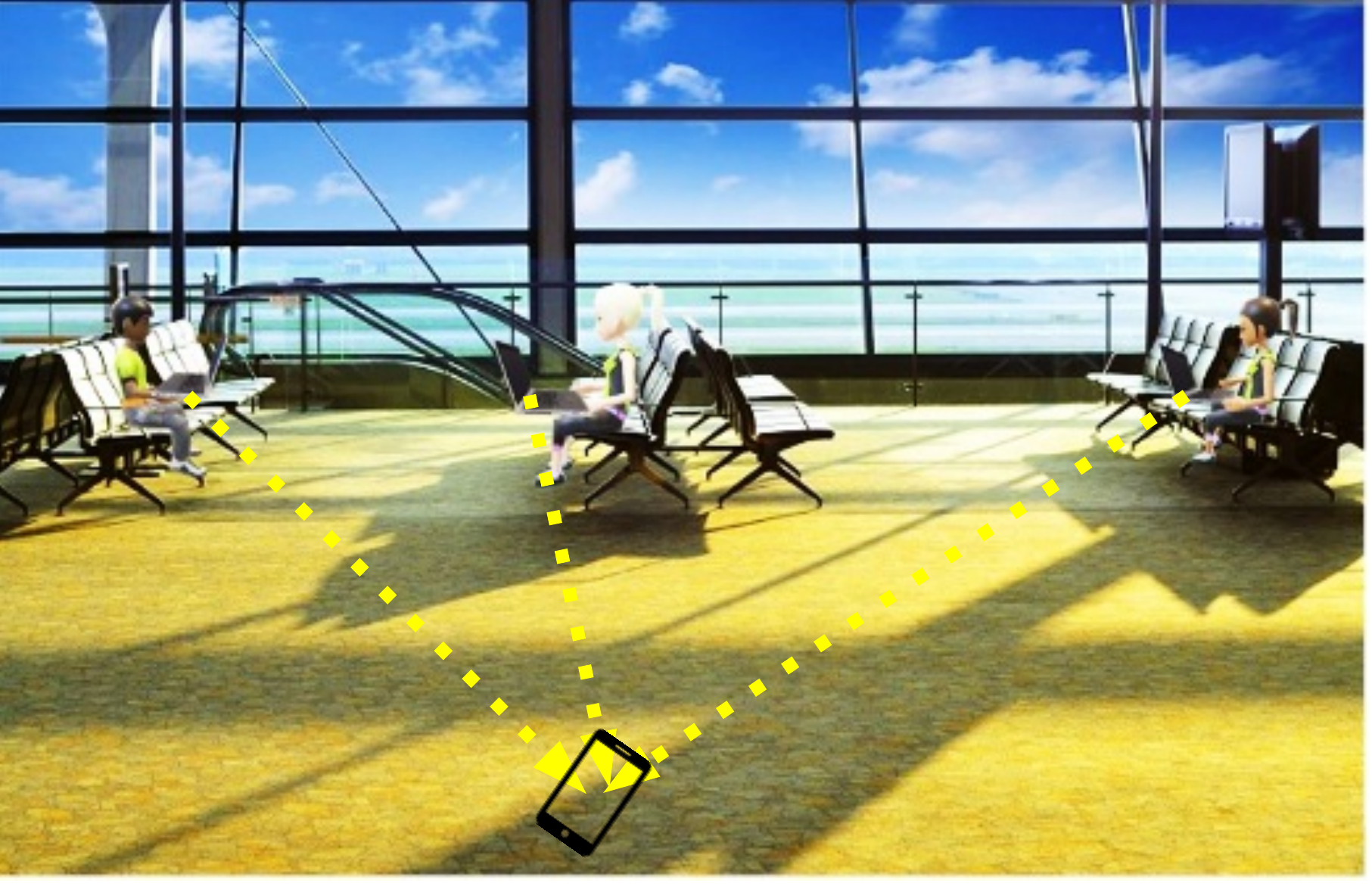}
\captionsetup{labelfont=bf}
\vspace*{2mm}
\caption{\textbf{\sysname}. Localizing network built in public places.}
\label{fig:scenario}
\end{figure}

Particularly, we design a system on the ubiquitous laptop platform with existing two antennas that doesn't require any additional hardware modification or external antennas, realizing accurate AOA based localization. As a result, the control of the localizing system is transfered from network administrator to  the public. As such, our system creates a new scenario where the localizing network is dynamically and collaboratively constructed by normal users. Therefore, with its accuracy and dynamic and distributed coverage, \sysname\ can enable a bunch of applications. 

In individual study room of a library, the steal of personal stuff is  a big concern to the students. With \sysname\ they can register their devices to the localizing system so as to  protect against stealing by detecting abnormal displacement. Differing from existing designs, \sysname\ system is dynamically constructed. No additional infrastructure is required and users can build their localization network at anywhere even with no wall power. Further, the distributional feature of freely-joined nodes in this system offers a better coverage and fair service to each device than static AP infrastructure. Another significant use case is in the airport (Figure \ref{fig:scenario}) or large public occasions.  It has been reported that in America every day more than 2000 children get lost in public places \cite{parentguidenews}. We envision \sysname\ can be an effort to remedy the situation. By running \sysname\ on their devices, people can  freely construct a localizing network in any public place even without infrastructure like routers or base station. Then, the lost children could be effectively and continuously tracked with a device like cellphone in their pocket or simply lost-and-found searching. Last but not least, \sysname\ could also enable activity organizer to promote interactivity among participants/audiences with this collaborative localizing system.
%, providing unique flexibility and accuracy for tracking the target  and have a good distributed coverage
%preventing children getting lost: more than 2000 children lost every day in public places in America (news http://www.parentguidenews.com/Articles/PreventingKidsFromGettingReallyLost)
%taking a picture in airport to show the application scenario https://securityinsights.pelco.com/locating-lost-children-in-airports

To design the above system, we need to tackle several challenges. First, the platform of interests has only two antennas which has lower resolution on AOA estimation compared with three antennas case. How can we achieve comparable accuracy despite of less number of antennas? Next, as we will see later, there is ambiguity in the AOA estimation brought by large antenna separation which is critical in localizing the target. However can we resolve this ambiguity? Furthermore, as to constructing a dynamic localizing network with freely joined devices (network nodes)  without any infrastructure, one key question is how to determine positions of those nodes which serve as the reference for localizing the target. That also means, \sysname\ enables a  maintenance and operation free service for the public. Unlike previous works, \sysname\ doesn't rely on any pre-measured anchor locations (e.g., SpotFi\cite{kotaru2015spotfi}) or opportunistic GPS service (e.g., \cite{chintalapudi2010indoor}). 

In \sysname, we propose our solutions for each of the above issues to enable the whole system. Firstly, we rely on an important observation that large separation of antennas (e.g., 26 cm on HP Pavilion) despite of the ambiguity issue could actually be utilized to solve the shortcoming of less number of antennas. State-of-the-art AOA based CoTS WiFi systems\cite{gjengset2014phaser,kotaru2015spotfi} use outer antennas connected to chipset whose spacing is controlled to be within half wavelength so as to avoid ambiguity. As such, the limited number of antennas and its large separation have hindered researcher to explore AOA based localization on existing laptop platform. While, we observe that large antenna separation also make the steering beam narrow, although there are ambiguous grating lobes. \sysname\ exploit this opportunity and process with improved music algorithm to enhance the number of effective sensors, boosting the accuracy of AOA estimation.

\begin{figure}[!t]
\centering
\begin{minipage}[t]{0.18\textwidth}
\subfigure[$\lambda/2$ separation]{
\includegraphics[width=\textwidth]{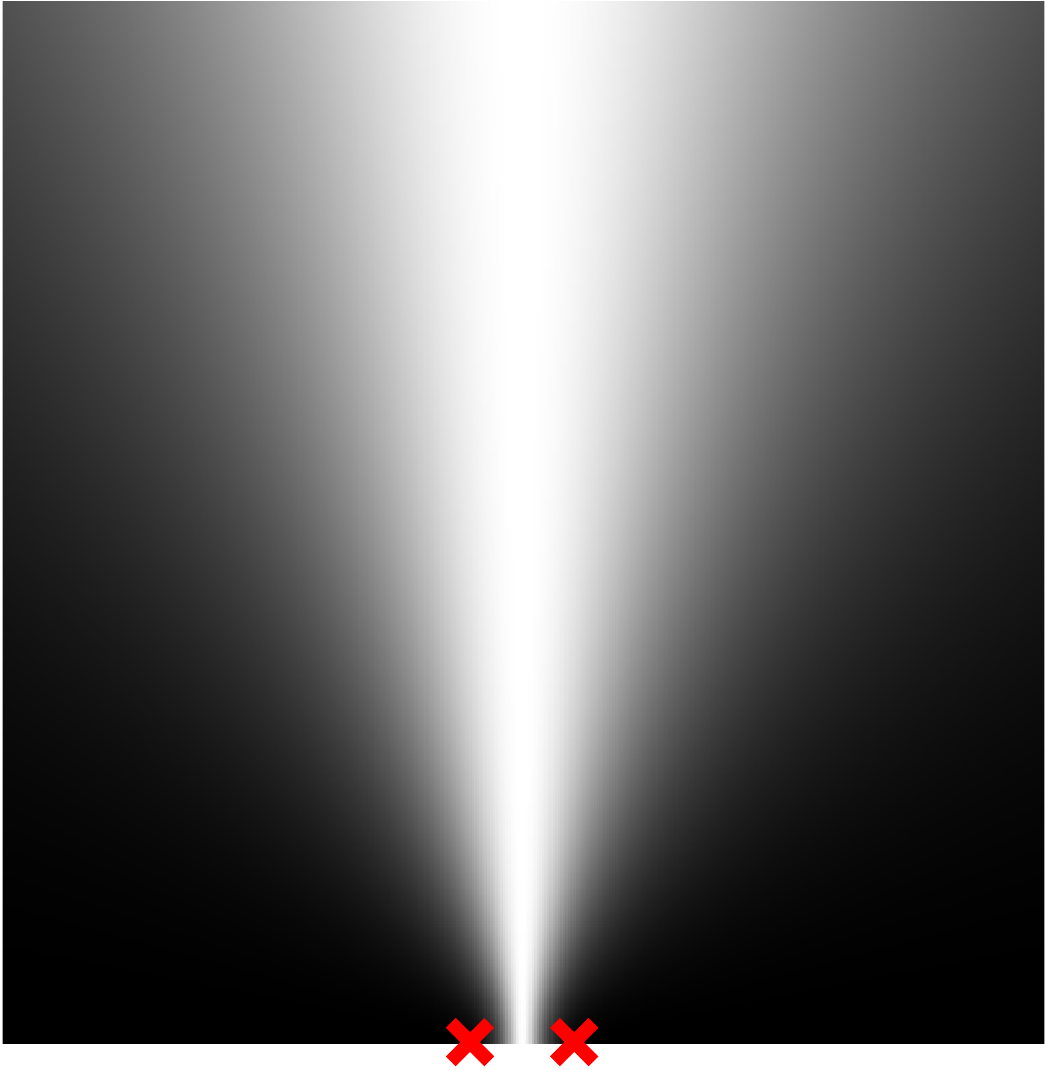}
}
\end{minipage}
\begin{minipage}[t]{0.18\textwidth}
\subfigure[$5\lambda/2$ separation]{
\includegraphics[width=\textwidth]{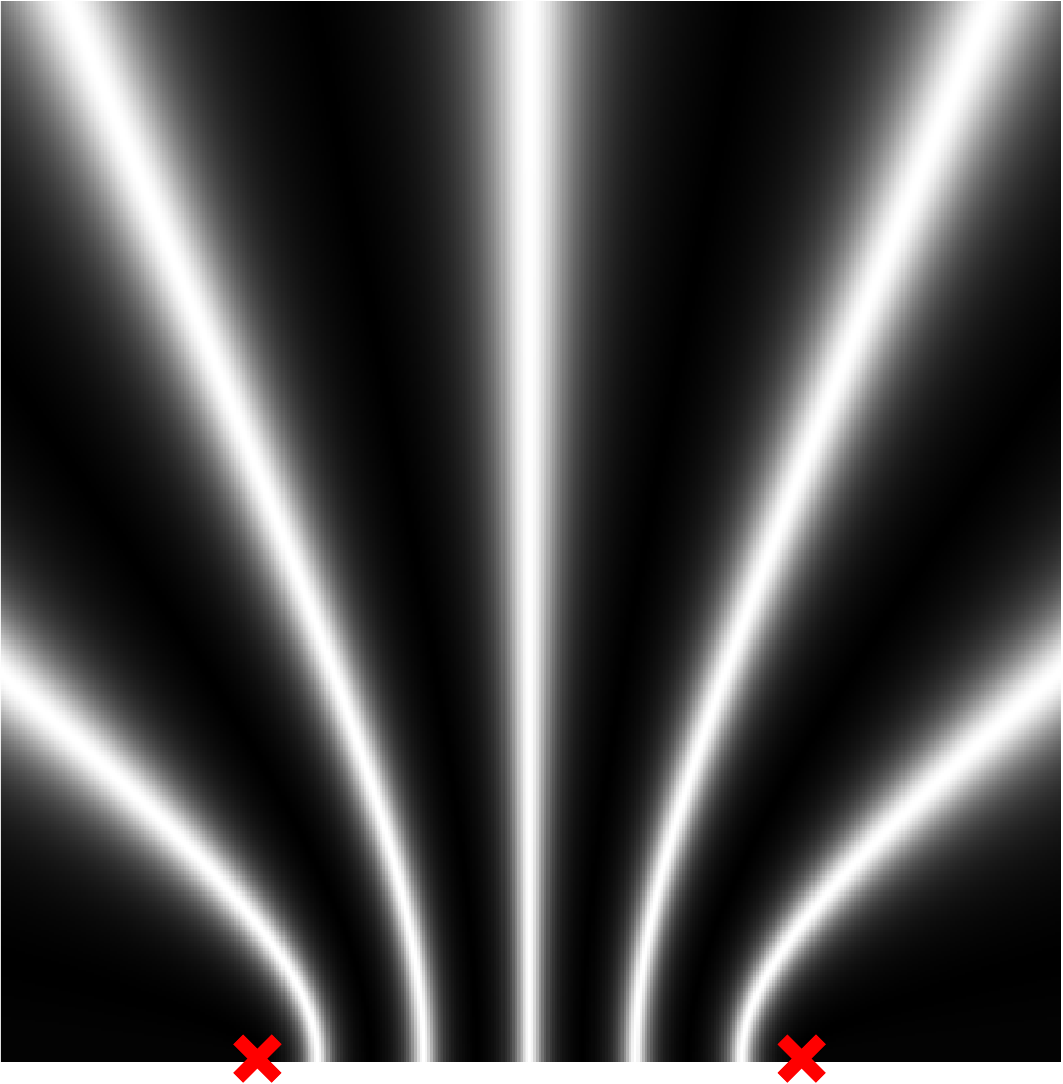}
}
\end{minipage}

\captionsetup{labelfont=bf}
\caption{\textbf{Beam pattern w.r.t antenna separation.} With antenna separation larger than $\lambda/2$, the steering beam becomes narrower but also confusing with grating lobes.}
\label{fig:beampattern}
\end{figure}

However, what comes together with the high resolution is ambiguity issue. Specifically, when antenna separation in an array is larger than half wavelength, the  beam pattern which is used to indicate the direction of signal source exhibits ambiguous grating lobes, shown in Figure \ref{fig:beampattern}. In \sysname, we design a multi-stage algorithm calculating the likelihood of candidates derived from AOA distribution and RSSI profile. By doing so, the ambiguity is resolved by selecting the candidate position with highest likelihood in the end. 

Last but not least, it is critical to localize the network nodes themselves which serves as reference for pinpointing the target by integrating their measurements. As stated before, \sysname\ is free of operation and maintenance through self-localizing. This is challenging since the network topology is dynamic with continuously joined devices and we do not require anchor APs\cite{joshi2013pinpoint} for authentic locations. In \sysname, we propose a non-intuitive algorithm called bundled two-layer self-localizing (BTS) which separates the  self-localizing problem into two layers, graphical topology description and scale size determination.  In this way, the algorithm is promising to separate the interference from these two layers and improve the overall accuracy.
% To evaluate our system, we run extensive experiments in typical class room environments.

We have built our system on ubiquitous laptops with minimum configuration of two antennas. Our system doesn't have any requirements or constraints on the signal source, which could be any WiFi device, such as laptop, phone, tablet or wearable device. In conclusion, in this work we make the following contributions:
\begin{itemize}
\item
We are the first to explore accurate AOA based localizing system design which can be controlled by normal users with their possessed ubiquitous devices without additional infrastructure. In particular, we design and implement our system upon normal laptops with minimum configuration of two antennas. 
\item
To achieve high accuracy with small number of antennas, our system exploits the narrow beams with large antenna separation on laptop. By resolving the ambiguity from multiple grating lobes, our system achieves comparable localization accuracy with state-of-the-art design \cite{kotaru2015spotfi} despite of less number of antennas. 
\item
To the best of our knowledge, \sysname\ is the first AOA-based dynamic and collaborative localizing system constructed  with no infrastructure, free of operation and maintenance. To achieve this feature, we design a dedicated two-layer algorithm which localizes the positions of network nodes with no anchor APs or opportunistic GPS service.
\item
We implement our design on ubiquitous laptops with minimum configuration of two antennas. To evaluate our system, we conduct extensive experiments on various channel conditions. As the results of experiments, our system shows median localization accuracy of 0.47 m in LOS scenario and 0.84 m in nLOS case. With self-localization, our system can still achieve median accuracy of 0.82 m in LOS scenario.
\end{itemize}

%% file: sec_Related.tex
\section{Related Work}
\label{sec:related}
%\vspace*{-0.3cm}
%Commodity device localization.
%WiFi adapter localization.
%RSSI based/ CSI based ( machine learning and finger-print/pattern based) not general usable/reliable.
%AOA based solution ( not accurate due to limited number of antennas).

CoTS WiFi based localization has primarily relied on RSSI and CSI information reported by commodity WiFi cards. RSSI based approaches\cite{bahl200radar0,chintalapudi2010indoor,wu2012fila,lim2005zero} measures the RSSI from the target at multiple APs. Mapping RSSI to distance via signal propagation model, those methods derive the location of the target with triangular relationship with intersection. Fingerprinting based approaches \cite{nandakumar2012centaur,azizyan2009surroundsense,yang2012locating,liu2012push} record the unique distribution of RSSI from one location to multiple receiving APs and match the pattern to work out the position. Similarly, \cite{wang2016lifs} analyzes patterns in CSI traces affected by multipath environments. Those approaches are subjected to the varying transmitter location and changing environments, which would incur expensive and recurrent operation to learn a matching configuration. In contrast, \sysname\ is based on angle of arrival (AOA) of RF signal, which maintains its confidence with varying surrounding environments. 

Prior AOA based CoTS WiFi localizing systems \cite{gjengset2014phaser,kotaru2015spotfi} has mainly relies on network infrastructure (e.g., AP router) with at least three outer antennas separated within half wavelength. The reason for constraining distance is to avoid ambiguity since multiple beams would occur as the distance exceeds half wavelength. In contrast, \sysname\ explores designing localization system utilizing normal WiFi device (i.e., laptops), which provide the opportunity to give back the control of localization system to normal users. To achieve high accuracy despite of the hardware limitation, we propose solutions to tackle the challenges arisen from limited number of antennas and large separations on the platform of interests. 

Within a localizing system,  the target is pinpointed by integrating measurements (i.e. distance or orientation) from multiple APs. Therefore, the locations of the network nodes (APs) should be known in advance to serve as reference for the target signal source. Differing from previous static network infrastructures \cite{lim2005zero,kotaru2015spotfi,bahl2000enhancements,sen2013avoiding,joshi2013pinpoint,niculescu2004vor}, \sysname\ is a dynamic localizing system constructed and expanded by freely joined users' devices, and hence need to work out the network spatial topology  by itself on the go. Previous works \cite{chintalapudi2010indoor,joshi2013pinpoint,vcapkun2002gps}  explored the problem of system self-localizing. However, they either require GPS service, which is not available in indoor places, to provide locations for several anchors or relied on coarse-measured distance derived from time-of-flight (ToF). By contrast, \sysname\ utilizes AOA profile targeting at accurate localizing performance without any anchor locations or GPS service.
%
%Dynamic localization network::: RSSI based localization network. AOA based localization network. 

In the context of RF-signal based positioning technologies, researchers have more freedom to design and implement systems on modified hardware\cite{xiong2013arraytrack,wang2014rf,gjengset2014phaser}, complex physical-layer signal engineering\cite{ma20163d} and large-scale/dense infrastructure\cite{sun2015widraw}. In contrast, \sysname\ is built on ubiquitous CoTS WiFi platform, which doesn't require any additional hardware or modified RF signal. Further, \sysname\ doesn't incur any cost if the CSI information is extracted from normal communication traffic. 

%various platform, software defined radio, modified hardware, complex signal processing in physical layer, large infrastructure. ... 

Non-RF based approaches have also been explored in the literature. \cite{shangguan2017enabling,wang2013rf,wang2013dude,shangguan2015relative} is RFID based systems which utilize radio reflection from tags attached on objects and process the signal on dedicated reader. Visible light based systems\cite{zhu2017enabling,hu2013pharos,kuo2014luxapose,zhang2016litell} need dedicated infrastructure for deployment and vulnerable to ambient light. Audio based systems\cite{tung2015echotag} are unreliable when exposed to the environmental noise in public places. In contrast, \sysname\ works on ubiquitous user-owned WiFi devices which enable a dynamic and collaborative localizing system with RF signal.

%% file: sec_Overview.tex
%\vspace*{-0.1cm}
\section{System Overview}
\label{sec:overview}
%\vspace*{-0.2cm}
%\sysname\ relies on AOA estimation to enable an accurate localizing system with distributed and dynamic networking nodes. 

To realize a robust localization system against changing environments, \sysname\ relies on angle of arrival (AOA) of the receiving signal from the object. Thus, it is helpful to understand how traditional AOA estimation  works with the well-known music algorithm. After that, we will introduce the system architecture to understand how \sysname\ works in real world and what are the components in our system design.
\subsection{Premier in music algorithm}
	 In the literature, music algorithm is  well-known and exploited  to extract AOA from the receiving signal. 
%	 Here we introduce the principle of music algorithm.
\begin{figure}[t!]
\centering
%\vspace*{-4.0cm}
\includegraphics[width=0.4\textwidth]{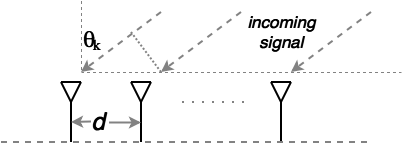}
\captionsetup{labelfont=bf}
\vspace*{2mm}
\caption{Arriving signal at the antenna array with AOA $\theta_k$.}
\label{fig:attarr}
\end{figure}
	The basic idea behind AOA estimation using music algorithm is to exploit the phase shift pattern\cite{orfanidis2002electromagnetic} of receiving signal across all the antennas of the array. When the target's signal arrives at the antenna array, the phase of the signal received at one antenna is closely related to the distance along the path it travels. Let $\Delta l_k$ be the additional distance traveled by the signal along the $k^{th}$ propagation path between two antennas. Then the phase shift is given by $2\pi \Delta l_k/\lambda$, where $\lambda$ is the wavelength of the arriving signal. Let us denote the complex exponential of the phase shift as a function of the distance along the path, i.e.,
	\begin{equation}
	\label{eq:eq1}
	\Phi(l_k)=exp^{-j2\pi \Delta l_k/\lambda}
	\end{equation}
	
	When the signal source is from some distance, the signals received at all the antennas in the array can be viewed as parallel and the additional distance traveled by the signal across consecutive antennas can be approximated by the antenna separation and the angle of arrival. As shown in Figure \ref{fig:attarr}, for the arriving signal at the angle of $\theta_k$ with respect to the normal of the antenna array, the additional distance traveled by the signal is given by $d\sin(\theta_k)$ where $d$ is the antenna spacing. Thus, we have the complex exponential of the phase shift as a function of the AOA of the path.
	\begin{equation}
	\label{eq:eq2}
	\Phi(\theta_k)=exp^{-j2\pi d\sin(\theta_k)/\lambda}
	\end{equation}
	Hence, with a uniform linear array with equal spacing of $d$ between consecutive antennas, AOA, i.e., $\theta_k$,  introduces a steering vector $s(\theta_k)$ of complex exponentials in the receiving signals across all the antennas in the array, i.e.,
	\begin{equation}
	\label{eq:eq33}
	s(\theta_k)=[1,\Phi(\theta_k)^{(1)},\Phi(\theta_k)^{(2)},...,\Phi(\theta_k)^{(M-1)}]^T
	\end{equation}
	where $M$ is the number of antennas in the array. For the signals received at all different antennas along one single path, the complex attenuations are same expect for the additional phase shifts and is equal to the received signal on the first antennas, denoted as $\gamma_k$. Hence, considering all signal paths in the environment from the source to the antenna array, the overall received signal on the antenna array is the superposition of the signals due to all paths, i.e.,
	\begin{equation}
	\label{eq:eq4}
	x= A\Gamma
	\end{equation}
	where $A=[s(\theta_1),s(\theta_2),...,s(\theta_L)]$ is the steering matrix combining all propagation paths and $\Gamma=[\gamma_1,\gamma_2,...,\gamma_L]^T$ is the vector of complex attenuations for each of the propagation paths. Note that our goal is to obtain the steering vector in matrix $A$, which is easy to derive the AOA of the propagation path, from the receiving signal $x$. To achieve this, MUSIC algorithm utilizes the property that the signal subspace in the autocorrelation matrix, i.e., $xx^H$, is orthogonal to the noise subspace. Thus, MUSIC algorithm proceeds by first calculating the noise space formed by eigenvectors corresponding to those minimum eigenvalues and then constructs the estimator function based on the orthogonality, which exhibits peaks on the AOAs of arriving signals. We omit the mathematical detail for brevity here but refer to the broad literature discussing this processing \cite{schmidt1986multiple,cheney2001linear,odendaal1994two}
\begin{figure*}[!t]
\centering
\begin{minipage}[t]{0.3\textwidth}

\subfigure[Single subcarrier]{
\label{fig:exppattern1}
\includegraphics[width=\textwidth]{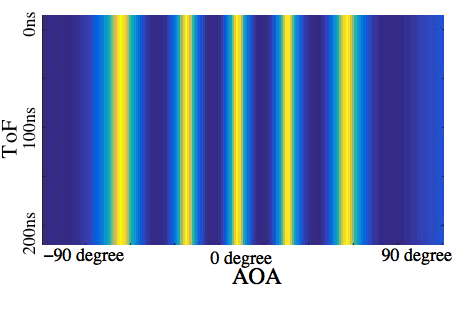}
}
\end{minipage}
\begin{minipage}[t]{0.3\textwidth}

\subfigure[Single eigenvalue]{
\label{fig:exppattern2}
\includegraphics[width=\textwidth]{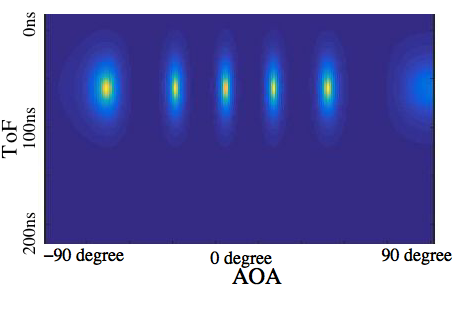}
}
\end{minipage}
\begin{minipage}[t]{0.3\textwidth}

\subfigure[All eigenvalues]{
\label{fig:exppattern3}
\includegraphics[width=\textwidth]{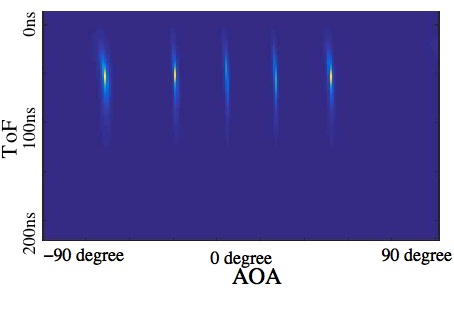}
}
\end{minipage}
\captionsetup{labelfont=bf}
\caption{\textbf{Beam pattern w.r.t number of subcarriers and eigenvalues.} With CSI values reported by laptop WiFi card on experimental LOS environment, we extract AOA profiles using  single subcarrier (Fig. \ref{fig:exppattern1}) and multiple subcarriers with single non-zero eigenvalue (Fig. \ref{fig:exppattern2}) and all 38 non-zero eigenvalues (Fig. \ref{fig:exppattern3}).}
\label{fig:exppattern}
\end{figure*}
\subsection{System architecture}
In this part, we will introduce the deployment platform of \sysname\ in real application scenario. Then, we briefly introduce three components in \sysname\ system design before we go to the detail of each part. 
\subsubsection{System deployment platform}
\sysname\ is a dynamic and infrastructure free localizing system. In real deployment, there is no limitation on where and when users can register their device to expand the network. \sysname\ system can be built in any place where there are some other devices nearby connected to this localizing network. Hence, there are two groups of participants in \sysname\ network, network nodes that serve the localizing function and registered clients which are WiFi devices to be tracked in this network. 

Typically \sysname\ works in this way. When an user enters a large public space, e.g., conference room or airport (Fig.\ref{fig:scenario}), where he wants to keep his device tracked  to protect from losing or steal, he opens the software App to search for local \sysname\ system. Once it is found, the user may either connect his laptop to expand the system as network node or just register any other devices, such as smartphone, tablet or smartwatch, to be tracked in this system. Note that it is needed to design an incentive mechanism to let users be willing to expand this system. An intuitive solution is to design a bonus for network node contributors which could be either a free service for tracking his devices or interests charged from other users who just register as clients. Since it is not the focus of this paper, we leave the study of incentive mechanism to future discuss. 

As far as the localizing network with distributed nodes is concerned, one node should be selected as the central server to integrate all information from other  normal nodes. To balance the computation across those nodes and reduce communication traffic load, in the design we let normal nodes finish as much computation as they can locally. Specifically, in addition to deriving the AOA profile respective to one normal node from one target, it continues to extract all necessary information from the AOA profile locally and then send to the central server. Hence, it doesn't need to send the whole CSI traces and also the central server doesn't need to search through the AOA profile exhaustively.
%
%csi information, normal laptop platform, prevalent and ubiquitous in major manufacturers. 
%The feature of this hardware distribution (range?random?distributed?coverage?dynamic?)
%select one as the central server, data processing in the central server and the networking nodes. notation in the following section.

\subsubsection{Components in system design}
There are three components in signal processing and algorithm in \sysname$\colon$

\noindent\textbf{Obtain AOA Profile.}
With improved music algorithm in Sec.\ref{sec:resaoa}, \sysname\ extracts the AOA profile from the receiving signal on the antenna array. With commodity WiFi cards, the receiving signal vector is reported as CSI values per subcarriers per antenna. In this AOA profile, the angle of arrival of the propagation path can be derived by searching the peaks across the whole direction range in the space.

\noindent\textbf{Resolve Ambiguity and Pinpoint the Target.}
Due to large separation of antenna pair on laptop, the estimated AOA profile has ambiguity in the target's direction demonstrated in Fig.\ref{fig:beampattern}. By integrating profiles from network nodes, we resolve ambiguity and pinpoint the target with the most likely candidate. 

\noindent\textbf{Network Self-localization.}
To enable a dynamic and infrastructure free system, we need to localize the network nodes themselves without any references. We solve this unprecedented challenge by designing a dedicated two-layer algorithm integrating the data from all network nodes, which separates the whole problem into graphical topology description  and scale size determination. 

%In the next section, we would go to the detail of each components in \sysname\ system design. 
%   the three step in localization (aoa profile, disambiguity and localization, self-localization)
%   architecture figure, showing the interconnection of network nodes (laptops). application scenario. 

%% file: sec_SysDesign.tex
%\vspace*{-0.4cm}
\section{System Design}
\label{sec:design}
In this section, we will discuss the three components in the signal processing and algorithm of \sysname\ as stated above. 

%First, we introduce how \sysname\ get high resolution of AOA profile from the CSI values reported by commodity WiFi card. Next, we talk about \sysname 's strategy in resolving the ambiguity in the AOA profile and pinpoint the target from multiple candidates. Lastly, we discuss how our system enables maintenance and operation free service for a dynamic network by self-localizing the network nodes themselves. 

\subsection{High resolution AOA profile}
 \label{sec:resaoa}
 To enable a distributed and dynamic localizing system free of infrastructure, \sysname\ needs to rely on the prevalent and ubiquitous personal owned WiFi devices - particularly laptops here. The limited number of antennas and large separation of its spacing hinders researchers to explore accurate AOA estimation on this platform. 
 Instead, in \sysname\ we solve this problem with the observation that although large separation of antenna pair would cause ambiguity in positioning, it also provide high resolution in AOA profile across the space. Besides, we utilize improved music algorithm to achieve a larger number of sensors despite of the fact that there is only two antennas. 
 
 To gain insight into the effect of antenna separation on the beam pattern, let's look at the Eq.\ref{eq:eq2} which is the approximated phase shift represented by AOA. For antenna separation less or equal to half wavelength, each phase shift corresponds to an unique AOA ($\theta_k$) which means there is one beam in the whole space. However, when the antenna separation becomes larger, more than one AOAs can generate the same phase shift and all of them could be considered as the direction of receiving signal, and hence cause ambiguity. Using more accurate model Eq.\ref{eq:eq1}, we show the beam pattern of separations of half wavelength and 2.5 wavelengths respectively in Fig.\ref{fig:beampattern}. As we can see, with half wavelength separation, there is no ambiguity but the beam is very broad thus low resolution. While, although multiple beams exist with larger separation, each beam is much skinnier and thus has high resolution. Specifically, with two antennas separated on a laptop by 26 cm, the average beam width (measured by -3dB range similar to bandpass filter) for 2.4 GHz WiFi signal is about one fourth as large as that with half wavelength separation.
 
 Further, we exploit the improved music algorithm to combine CSIs on all subcarriers so as to provide better resolution of the AOA profile. The basic idea is to extend the number of effective sensors by incorporating time of flight into steering vector and smoothing CSI matrix. In previous literature \cite{vanderveen1998estimation,van1998joint,xiong2013arraytrack,paulraj1986performance,kotaru2015spotfi}, jointly estimation of AOA and TOF and CSI smoothing have been discussed. Here, in \sysname\ we need to figure out the corresponding design of algorithm based on the above ideas on the laptop platform with just two antennas. 
 
 Due to frequency difference, there is a phase shift between consecutive sub-carriers on the same antenna, given by $\Delta\varphi=2\pi(f_i-f_j)\tau_k$ where $f_i,f_j$ are the frequencies of these subcarriers and $\tau_k$ is the ToF of the $k^{th}$ propagation path. Let's denote the complex exponential as $\Psi(\tau_k)$. Hence, if we considering all CSI measures on $N$ subcarriers as opposed to just one in Eq.\ref{eq:eq4}, we can obtain the following,
	\begin{equation}
	\label{eq:eq5}
	X= AF
	\end{equation}
where $X=[x_1, x_2, ..., x_N]$ and $F=[\Gamma_1, \Gamma_2, ..., \Gamma_N]$. From theoretical analysis, the resolution of Eq.\ref{eq:eq5} on estimating AOAs is limited by the number of rows in matrix $A$ and the number of columns in matrix $F$. Note that the number of rows in matrix $A$ is the number of antennas in the array while the number of columns in matrix $F$ corresponds to the number of measurements, i.e., number of subcarriers. Particularly, in our platform of interests (laptop), the number of antennas is only two which is much less than the total number of subcarriers in the 20MHz WiFi band. Specifically, with the commodity Atheros chipset, the CSI matrix for each packet received would contain channel information for 56 subcarriers in total. 

To solve this unbalance so as to boost the number of effective sensors, we apply a mathematical trick as improved music algorithm. The basic idea is to rearrange the matrix formula equation in Eq.\ref{eq:eq5} so that the number of rows in matrix $A$ and the number of columns in matrix $F$ is balanced. Specifically, let's look at the rearranged formula in Eq.\ref{eq:eq6} to understand our design.
	\begin{equation}
	\label{eq:eq6}
	X^\prime= A^\prime F^\prime
	\end{equation}
where  $A^\prime=[s^\prime(\theta_1),s^\prime(\theta_2),...,s^\prime(\theta_L)]$ and $s^\prime(\theta_1)$ is the extended version of $s(\theta_1)$ in Eq.\ref{eq:eq33}. Recall that $s(\theta_k)=[1,\Phi(\theta_k)]^T$ since $M=2$. Here, we extend this vector with 19 consecutive subcarriers and obtain $s^\prime(\theta_k)=[1,\Psi(\tau_k),\Phi(\theta_k),\Psi(\tau_k)\Phi(\theta_k),...,\Phi(\theta_k)^{(18)},\Psi(\tau_k)^{(18)}\Phi(\theta_k)]^T$. Accordingly, the same number of consecutive CSI measurements on 19 subcarriers should be stacked in each column of extended matrix $X^\prime$, i.e., $x_1^\prime=[csi_{1,1},csi_{1,2},...\newline,csi_{2,18},csi_{2,19}]$ where $csi_{i,j}$ is the CSI measurement for $i^{th}$ antenna and $j^{th}$ subcarrier. 

Now, let's consider the columns of matrix $X^\prime$. As we want to achieve the linear combination in equation \ref{eq:eq6}, we shift each entry in the first column of $X^\prime$ by one subcarrier, i.e., $x_2^\prime=[csi_{1,2},csi_{1,3},...,csi_{2,20}]$. In this way, the linear combination in the above formula is maintained as long as we multiply a common scaling vector $[\Psi(\tau_1),\Psi(\tau_2),...,\Psi(\tau_L)]$ to the first column of $F^\prime$ to obtain its second column. As a result, the number of columns in matrix $F^\prime$ is 38 which is the same as the number of rows in matrix $A^\prime$.

In Fig.\ref{fig:exppattern}, we obtain CSI values from laptops in experimental LOS environment and calculate the AOA profiles in three ways. First, we just utilize one subcarrier on the pair of antennas for AOA calculation and see the variation across all subcarriers as shown in Fig.\ref{fig:exppattern1} which doesn't have ToF resolution. Next, we use CSI on all 56 subcarriers using the proposed improved music algorithm. To look into the effect of extended number of effective sensors, we calculate using just one (Fig.\ref{fig:exppattern2}) and all 38 (Fig.\ref{fig:exppattern3}) non-zero eigenvalues. Comparing all three results, we can see that the proposed music algorithm improves the resolution of AOA profile with a much skinner beam. To measure the localizing accuracy with the AOA profile, we further conduct experiment in Sec.\ref{sec:aoast} and obtain the median error as 2.25 degrees, comparable to the state-of-the-art localizing system with CoTS WiFi. 
 \begin{figure}[b!]
 \centering
 %\vspace*{-4.0cm}
 \includegraphics[width=0.4\textwidth]{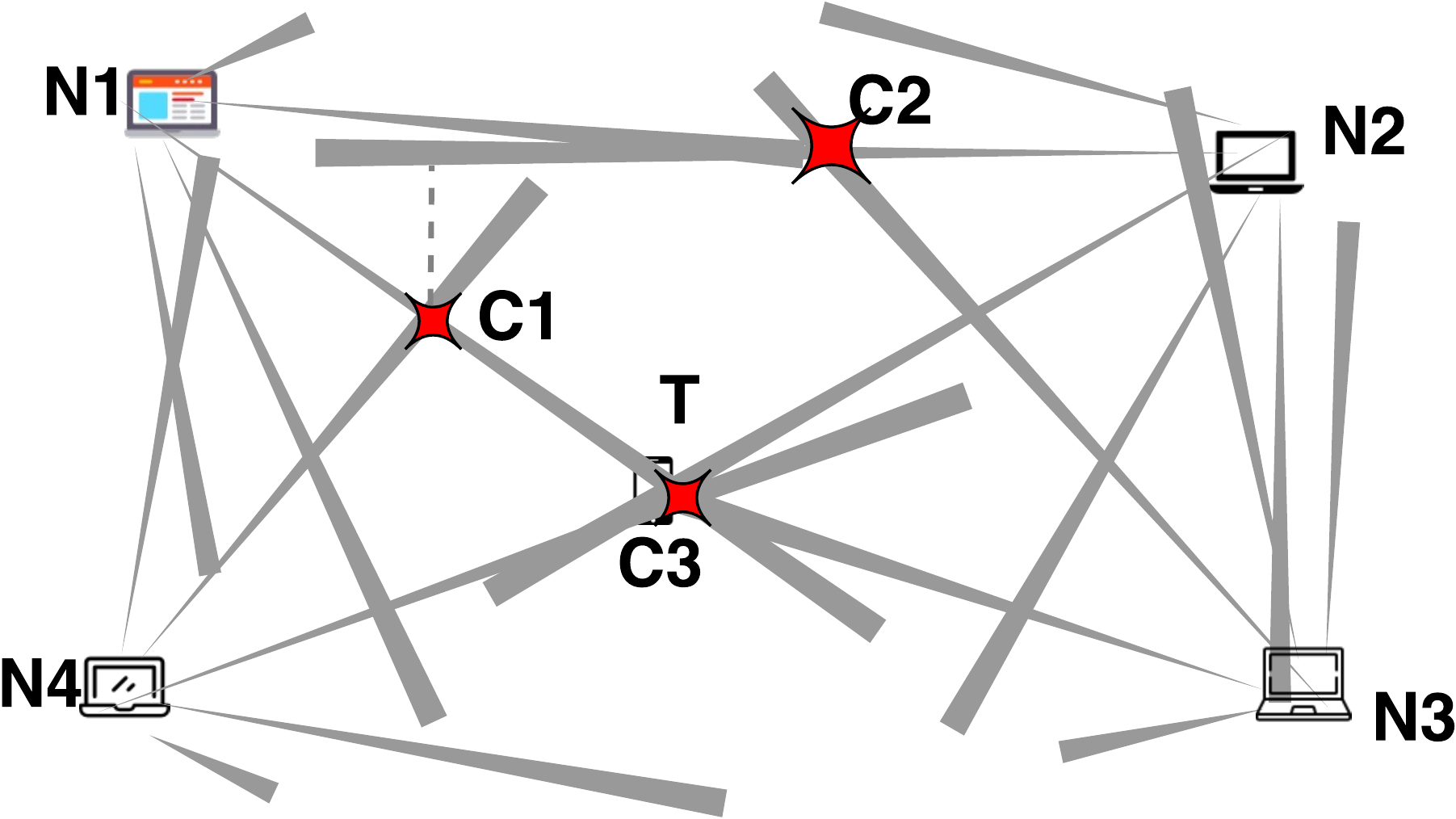}
 \captionsetup{labelfont=bf}
 \vspace*{2mm}
 \caption{Resolve ambiguity in localizing network of \sysname.}
 \label{fig:netloc}
 \end{figure}
   \begin{figure*}[t!]
   \centering
   \begin{minipage}[t]{0.24\textwidth}

   \subfigure[Error range based selection]{
   \label{fig:sim1}
   \includegraphics[width=\textwidth]{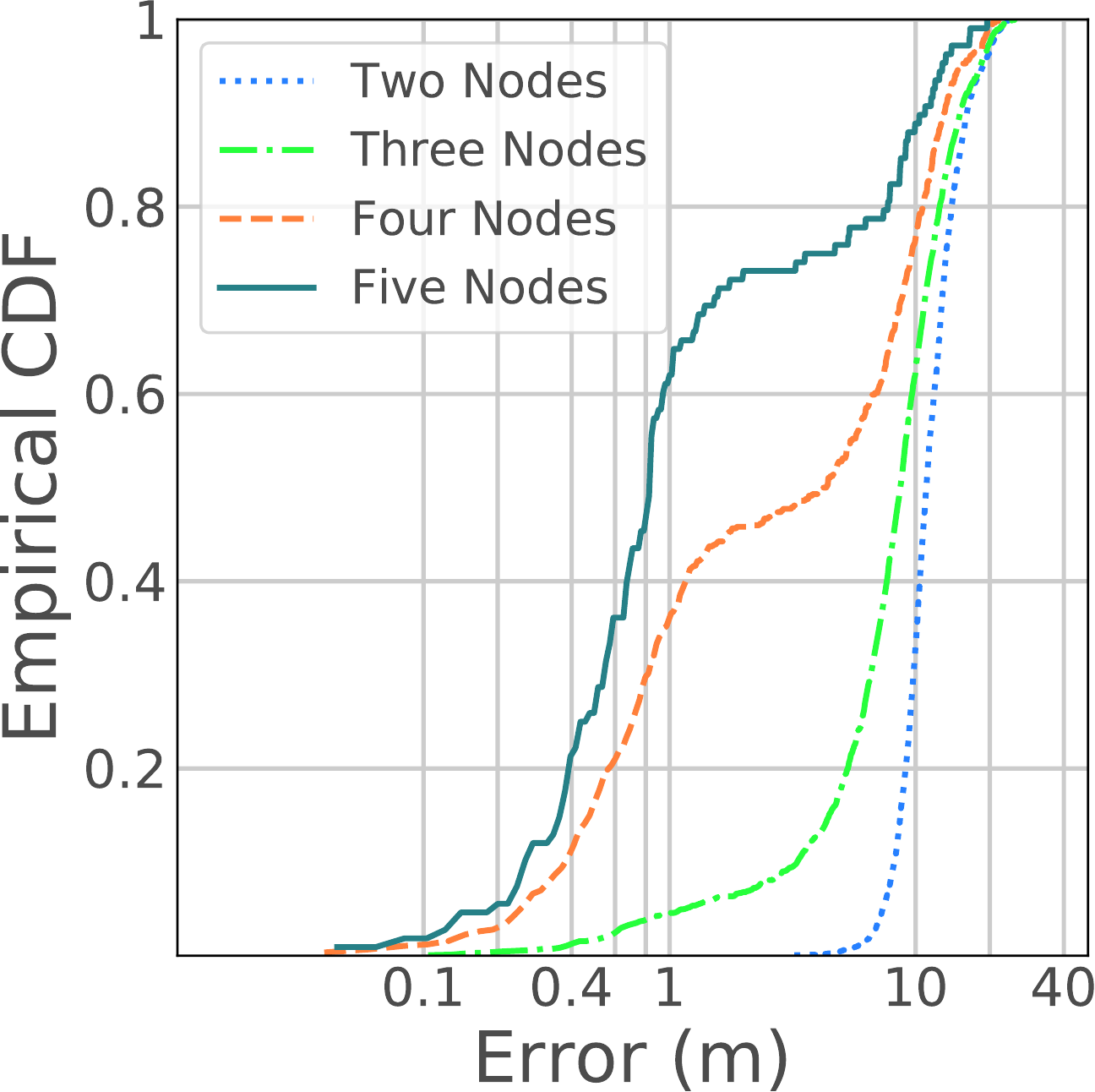}
   }
   \end{minipage}
   \begin{minipage}[t]{0.24\textwidth}

   \subfigure[Likelihood based selection]{
   \label{fig:sim2}
   \includegraphics[width=\textwidth]{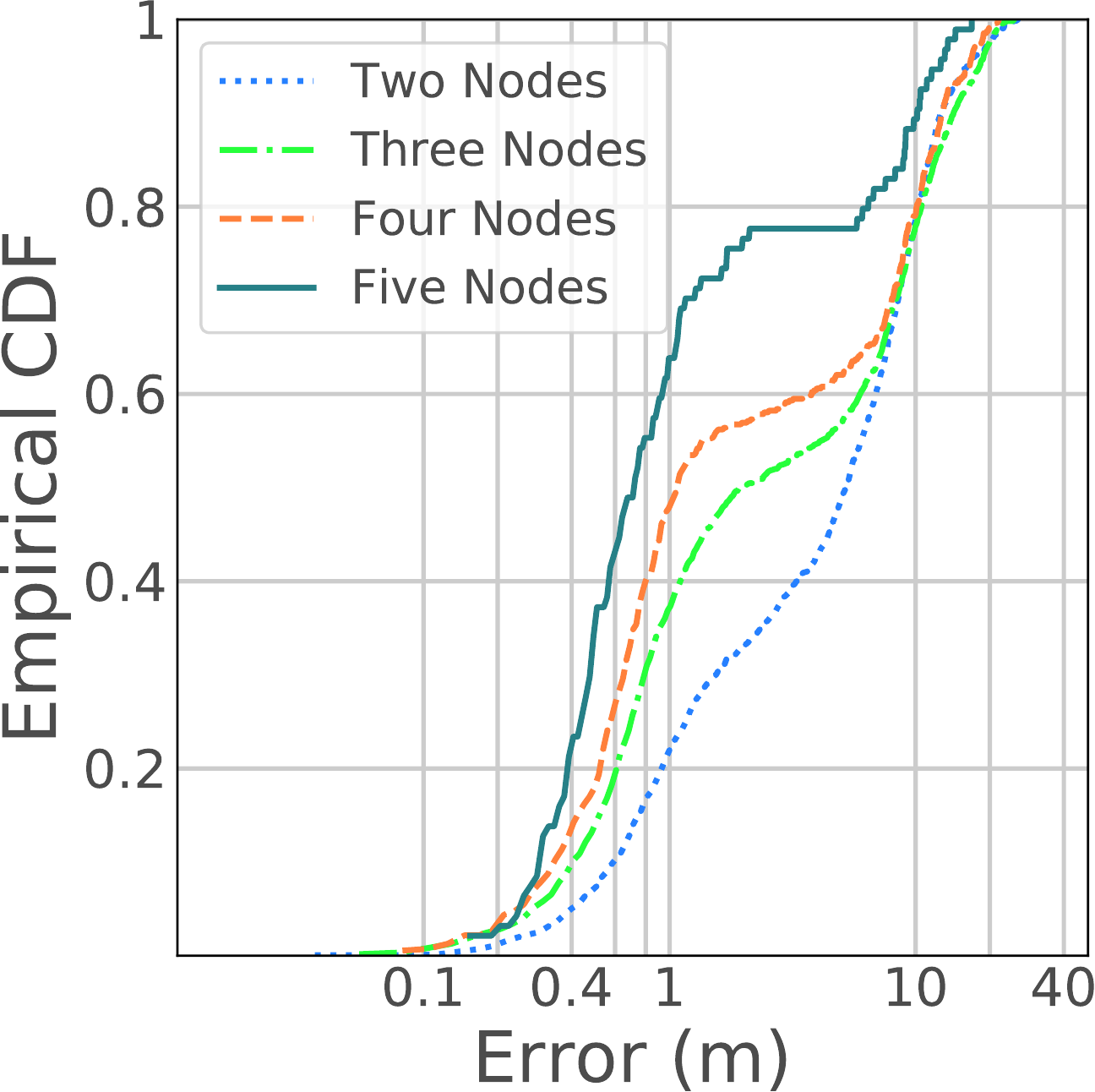}
   }
   \end{minipage}
   \begin{minipage}[t]{0.24\textwidth}

   \subfigure[Measurements integration]{
   \label{fig:sim3}
   \includegraphics[width=\textwidth]{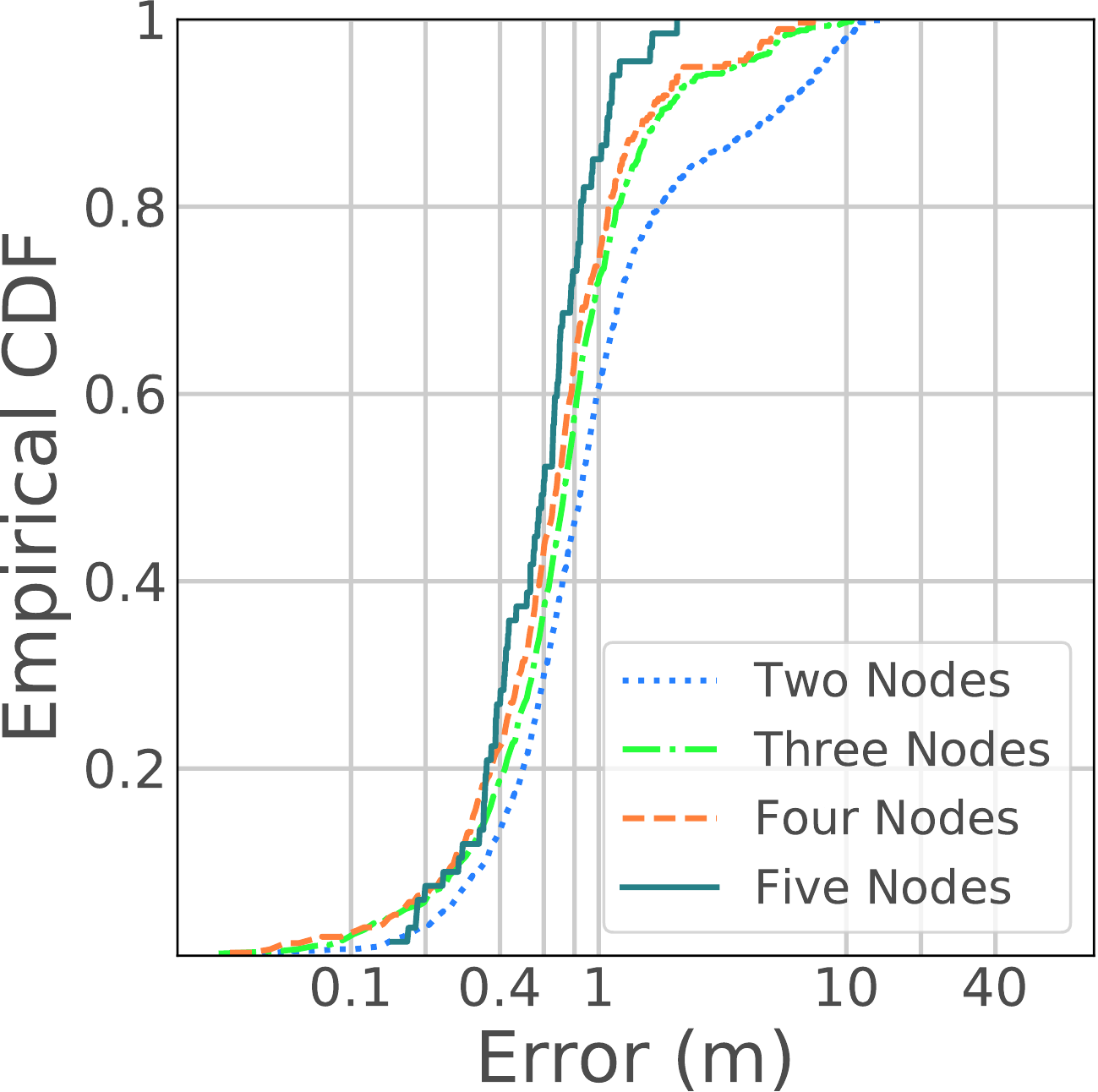}
   }
   \end{minipage}
   \begin{minipage}[t]{0.24\textwidth}

   \subfigure[Target localization]{
   \label{fig:sim4}
   \includegraphics[width=\textwidth]{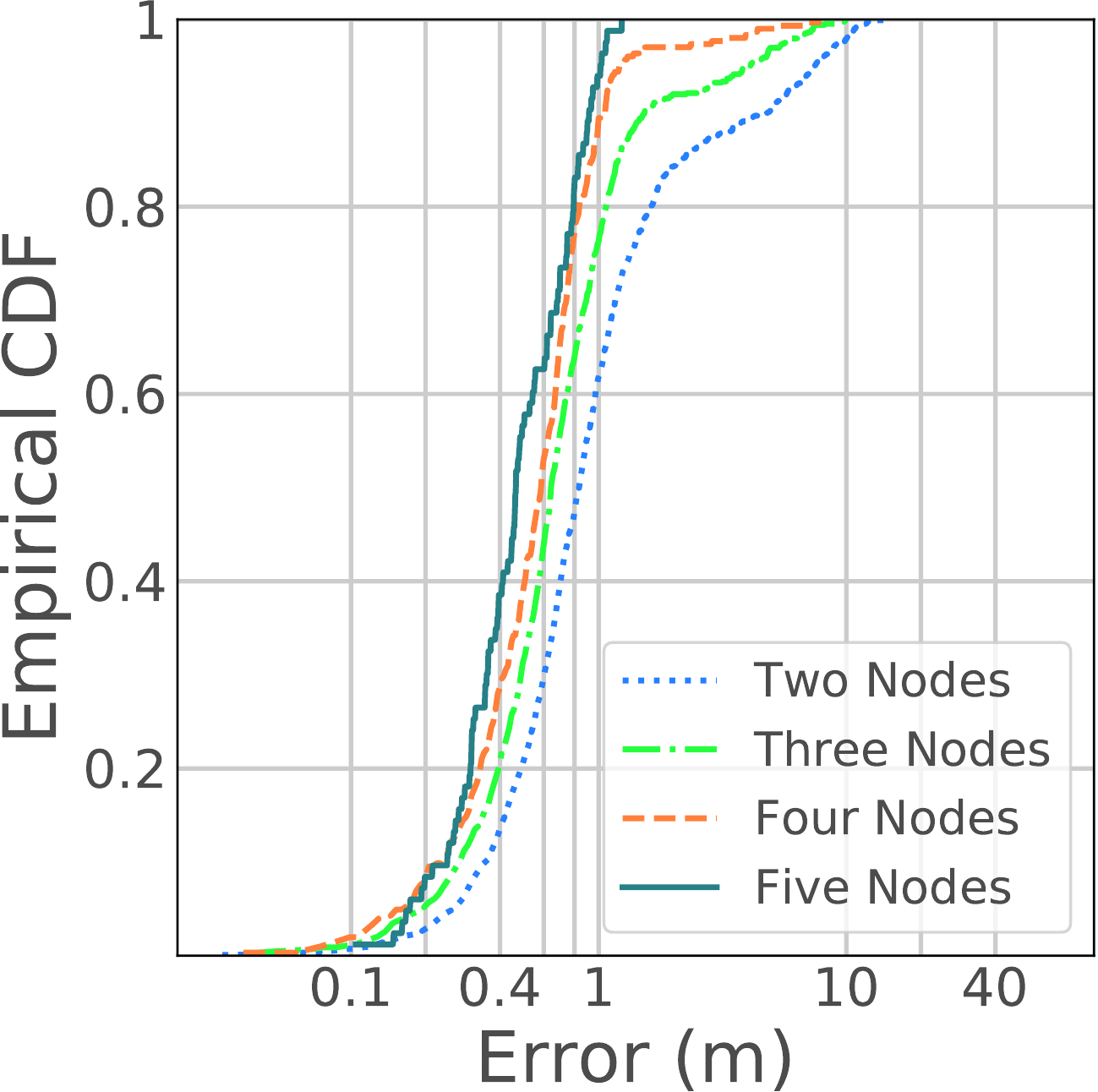}
   }
   \end{minipage}
   \captionsetup{labelfont=bf}
   \caption{\textbf{Simulation results on different stages of ambiguity resolution and target localization.} With simulation on a space of 30mx50m, we test on the effect of different factors in the ambiguity resolution and target localization, using the experiment result on AOA estimation in LOS scenario (Sec.\ref{fig:aoa1}) as the error distribution. These factors are separated into four stages as stated in Sec.\ref{Sec:amb}. We also look into the effect of network-node density by setting node number as 2, 3, 4, 5 respectively in the simulation.}
   \label{fig:simul}
   \end{figure*}
\noindent\textbf{Target in near scope.} The approximation in Eq.\ref{eq:eq2} using AOA is not accurate if the signal source is within near scope of the antenna array. Fig.\ref{fig:beampattern} demonstrates that each AOA actually corresponds to one hyperbola. Thus, in near scope we use more precise model represented by Eq.\ref{eq:eq1} to replace AOA model in the steering vector. 
% the purpose this this section
% the accurate aoa estimation combining super-resolution music algorithm and large-separation fine-grained grating lobes. 

\subsection{Ambiguity resolution and localization}
\label{Sec:amb}
 In the above, we obtain the high resolution of AOA profile  as the first step. However, unlike state-of-the-art AOA localizing works which directly intersect the unique beams to pinpoint the target due to small antenna separation, there is challenge coming from the ambiguity of additional lobes to get the correct location. In \sysname, we make use of multiple measures and resolve the ambiguity before pinpointing the target. 
 
 \noindent\textbf{1. Ambiguity resolution}
 
 In the first, our intuition in resolving the ambiguity on multiple grating lobes is the observation that the correct location would always be the intersection in the AOA profiles of any two network nodes. In contrast, for the counterfeit location, with more network nodes joined, it is more unlikely that it will satisfy the AOA profiles for all of them. Hence, as long as we integrate the AOA profiles from multiple network nodes up to some number, we should be able to resolve the ambiguity to pinpoint the only correct location. Let's look at the example shown in figure \ref{fig:netloc}. If we only count for two network nodes, candidates C1, C2 and C3 all satisfy. However, as we increase the number of network nodes, C2 matches the AOA profiles for three network nodes and only C3 matches the AOA profiles for all four network nodes. As such, we can conclude that with enough network nodes the correct candidate location could finally be found. 

 However, due to wireless noise and hardware imperfection, noisy CSI measurements would cause variation in the AOA profile. Taking this into consideration, we note that no location can perfectly satisfy all AOA profiles and those which are within the error range should be kept as potential candidates. Therefore, we first search for candidate cluster which has intersections from all network nodes and thus represent selecting one beam for every AOA profiles. In order to avoid searching across the whole space, we first generate the candidates using intersections of the first two AOA profiles. Starting from the first group of candidates, we incorporates intersections of other pair of AOA profiles and eliminate the one exceeding out of the error range.

 To have an insight into how the integration works w.r.t the number of network nodes, we conduct a simulation which randomly selects the locations of network nodes and the target. In this simulation, we set the range of the whole space as 30mx50m and the error range of AOA profile of each network node from experiment as $[-8^{\circ},8^{\circ}]$, which covers 90 percentage of all measurements as shown in AOA evaluation Sec.\ref{fig:aoa1}. We exclude measurements out of this range since large error would greatly impair localization accuracy. However, this doesn't hurt performance of our system since 90 percent measurements are retained. Specifically, we measure the error of one candidate cluster using the mean position of its intersections as the candidate position. In Fig.\ref{fig:sim1}, the results show that with more network nodes SpotFi has higher accuracy on cluster selection, i.e., determining AOA candidate on each AOA profiles. Specifically, the median error of localization with five nodes has improvement of 9 m over that with just two nodes. 
 
Can we do better? To improve the cluster selection accuracy with less number of network nodes, we design a likelihood measurement using RSSI profile across all network nodes to further resolve the ambiguity. In the literature, RSSI profile is also explored to track the RF target and the median accuracy is around 4m constraint by insufficient RSSI model in changing environments. Albeit its low resolution, there is opportunity of improving accuracy by matching RSSI profile in \sysname. The reason is the observation that the candidate locations could be far from each other as shown in Fig.\ref{fig:netloc} since the grating lobes are separated by large angles and as we know wireless signal strength typically degrades exponentially w.r.t the traveled distance with a exponent around 2-4. Here, the likelihood measurement is designed to be negatively related to the overall deviation across all $I$ network nodes for one candidate of one cluster.
 \begin{equation}
 \label{eq:eq7}
 \begin{aligned}
 li&kelihood_C=\sum_{i=1}^{I} (r_i^2+r_i^{\prime 2})/\sum_{i=1}^{I} (r_i-r_i^{\prime})^2
 \end{aligned}
 \end{equation}
 where $r_i$ is the measured RSSI for node $i$ and $r_i^{\prime}$ is the estimated RSSI from propagation model\cite{bahl200radar0,chintalapudi2010indoor,wu2012fila,lim2005zero}. $C$ is one of the candidate clusters. Specifically, the RF propagation model used here is the log-distance path loss (LDPL) model, which is reviewed below. 
 \begin{equation}
 \label{eq:eq9}
 r_i=R-10\gamma logd_{i}
 \end{equation}
 where $d_{i}$ is the distance between $i^{th}$ node and the target. $R$ is the RSSI observed at a distance of 1m away from the target and $r_{i}$ is the RSSI observed at the target. The path loss exponent $\gamma$ captures the rate of fall of signal strength around the target. Those unknown parameters are estimated in the initialization procedure explained in Sec.\ref{Sec:netself}. In Fig.\ref{fig:sim2}, the results show that by integrating the likelihood measurements localization error could be reduced. We observe that the most improvement is on the case with just two network nodes. The reason is that two network-node system has no opportunity to detect counterfeit cluster with just one group of intersections and thus benefits the most from RSSI profile matching. 
 
\textbf{Integration of Multiple Measurements.}
Since there is random variation coming from wireless channel and hardware,  single packet would experience large deviation in the WiFi card measurements. In this case,  either wrong cluster would be chosen or large deviation exists in current measurement, resulting in large error. Fortunately, we found that the likelihood value is a very good indication on the quality of current measurement. For example, for network of 3 nodes, candidate position with error less than 1 m typically has likelihood value larger than 20 while those with error larger than 10 m has likelihood less than 1. Thus, for integration of multiple measurements, we reject bad measurements with a likelihood threshold to improve overall accuracy. Fig.\ref{fig:sim3} shows the results after integrating multiple measurements from 10 packets. We can see that number of large error cases in the cumulative distribution is significantly reduced and the media error for two-node and three-node case is reduced by more than 2 m. 
 
 \noindent\textbf{2. Localizing the target}
 
 Recall that in ambiguity resolution we obtain the candidate location by averaging all intersections in one cluster. The candidate location is used for measuring the accuracy of corresponding cluster and integrating RSSI profile. However, it may not be the perfect location even if the cluster is the correct one. Intuitively, the optimal location of the target should minimize the deviation across all AOA profiles. Besides, RSSI profile could also be utilized to measure the quality of candidate location. Based on these ideas, we design a likelihood measurement to be  negatively related to the overall deviation from all AOA profiles and mismatch between measured and  estimated RSSI profile for one candidate location $p$, which is below.

\begin{equation}
\label{eq:eq8}
\begin{aligned}
&likelihood_p=1/\\
&(\sum_{i=1}^{I} (r_i-r_i^{\prime})^2/\sum_{i=1}^{I} (r_i^2+r_i^{\prime 2})+\omega(I)\sum_{i=1}^{I}\Delta AOA_{i})
\end{aligned}
\end{equation}
where $\Delta AOA_{i}$ is the AOA deviation from candidate location $p$ to the AOA profile of $i^{th}$ network node. Then, with searching through all candidate locations in the scope of selected cluster, we pinpoint the target with largest likelihood. Note that while taking into account the error range in AOA profile, we set the searching scope to an expanded range of the cluster formed by all its intersections. Fig.\ref{fig:sim4} shows the localization accuracy with fine-grained searching in the cluster area, where the media error with five nodes is improved by 0.15 m. 

Compared to previous work \cite{kotaru2015spotfi}, our design of localizing algorithm has an additional stage to solve the ambiguity arisen from multiple grating lobes. In designing the ambiguity resolution algorithm, we make use of the error range measured from experiments in Sec.\ref{sec:aoast} to shrink the search space and RSSI profile to improve the performance of ambiguity resolution. Besides, due to the dynamic feature of \sysname, we use weight factor $\omega(I)$ to account for different scale of the network, i.e., $\omega(I)$ is an decreasing function of the number of network nodes. Further, in our design, the likelihood value is incorporated into integration of multiple measurements across consecutive packets to mitigate the random variation. 
% Fig. shows how the algorithm works with the additional likelihood measurements with the same simulation setting. Compared to the results in Fig., we can see that the improvements.... 
% 

% ..In fig(4), the updated performance of the final algorithm shows that with 2 network nodes the average number of candidates while with 3 network nodes the number decreases to ... This is a favorable result since in public places the number of network nodes would easily exceeds 5 or 10. 
% motivaiton of the design from observations in example candidates, error ranges, RSSI profile, increasing number of network nodes.
% the purspose of this section
% unique disambiguity algorithm
% the processing of intersection for final localization.
% combat the illegal one. malfunction or ..
% computation distribution: data processing in the networking nodes and in the central server.
% 

 \textbf{Practical Issue.} In practice, one issue of the above localizing algorithm is the non-line-of-sight (nLOS) scenario. Depending on the obstruction, there may or may not be direct path from the target to one network node. In the first case, we could get more accurate AOA whose power is not the strongest. In the second case, we need to suppress the AOA from the corresponding node since there is no direct path hence large error exists. We present detailed solution in Sec.\ref{sec:aoast}. 
\subsection{Network platform self-localizing}
\label{Sec:netself}
 In \sysname, we enable a distributed and dynamic localizing system free of infrastructure. As such, the normal user could register their own device to be a node in the network and also log out as they wish. Up to now, we have demonstrated how \sysname\ provides an accurate localization on the target using the AOAs relative to all network nodes. However, given the dynamic feature of this platform topology, the remaining question is how might \sysname\  figure out the localizations of the network nodes themselves. Our problem is different from previous RSSI-based self-localization works as we have accurate AOA profile to improve the accuracy. 
% Here, we design a two-step strategy in solving this problem. 
% the motivation and reason of this section.

 An intuitive and simple solution one might think of is to incrementally update the location of the new node, which we call incremental self-localizing update algorithm (ISU). Specifically, when a WiFi device is added to the current network as a new node by an user, ISU algorithm treats the new node as a client and obtain its location to update the network topology construction using the solution in Sec.\ref{sec:resaoa} and Sec.\ref{Sec:amb}. Despite its simplicity, we find there are two problems in ISU which deter it from working effectively in practice. First, as there exists deviation in the previous self-localization rounds, using the existing positions of network nodes to localize the new one would accumulate the deviation in sequence. What's more, such an algorithm doesn't fully utilize valuable sensing data from the new node. An important observation here is that the new node is also capable to build AOA profiles for existing nodes. 
 
 Taking these two issues into account, we design a new strategy called bundled two-layer self-localizing algorithm (BTS). The first part of the idea behind BTS is to bundle the updates of positions for all network nodes not just for the newly joined one. As such, it is promising to tackle the problem of accumulated error from the incremental operation of ISU. Besides, the second part is to separate the localizing problem into two layers. The intuition of this approach is that we find these two layers are quite independent in this problem and thus solving each of them separately may reduce mutual interference from the deviation of each other. In the following, we will go to the detail of the design in our algorithm. 
 
 \noindent\textbf{1. Graphical topology description}
 
 In the first layer, we utilize the linear relationship to effectively resolve the directions between any two network nodes and construct the graphical topology description for the whole network system. Specifically, different from just localizing the RF target, localizing of the network nodes has an unique advantage that two directions between two network nodes can be measured and they should match with each other. As shown in Fig.\ref{fig:netloc}, network nodes $N1$ and $N2$ could measure the directions of the signals emitted from each other and get two AOA profiles. It is easy to see that the ambiguity of AOA profile is effectively resolved since the two directions should be in the same line. Therefore, after obtaining the directions between any two network nodes, we construct a graphical description of the network topology. 
 
% minimize the deviation of over all aoa profiles.
% addition: we could utilize the triangle property to better improve the accuracy. That is, taking three nodes for example, the three angles reported as between any two-node pair should add up to 180 degrees. 
  
 \noindent\textbf{2. Scale size determination}
 
 After the computation in the first layer, although we only obtain the directions among network nodes, there is just one measurement remaining to be solved - the network scale size. The network scale size essentially describes how large space the whole network takes up. With the topology of network nodes, we could derive the ratio between any two lines. Mathematically, let's denote the unknown distance between $1^{th}$ and $2^{th}$ nodes as $d_{12}$ and between $i^{th}$ and $j^{th}$ nodes as $p_{ij}d_{12}$, where $p_{ij}$ is the ratio derived from the topology. With the log-distance path loss (LDPL) model described in Eq.\ref{eq:eq9}, we get the following equations.
\begin{equation}
\label{eq:eq10}
r_{ij}=R_i-10\gamma_ilog(p_{ij}d_{12})
\end{equation}
where $R_i$ is the RSSI observed at a distance of 1 m of $i^{th}$ node and $r_{ij}$ is the RSSI observed at $j^{th}$ node. Since $R_i$ is a device-specific metric, we assume that each device knows its own $R_i$. The path loss exponent $\gamma_i$ captures the rate of fall of signal strength around $i^{th}$ node.

In equation \ref{eq:eq10}, there are four parameters for each network node, which are $R_i$, $\gamma_i$ and two coordinates of $i^{th}$ node. Since we have $I$ network nodes, the total number of equations formed by the LDPL model is $I(I-1)$. Note that in order to determine the distance between each pair of nodes, we only need the scale size of the whole network.  Thus, the total number of unknowns is actually $I+1$. As long as $I > 2$, it is feasible to solve all unknowns. In the case that we just have two network nodes, note that even if there is just one client the condition is satisfied. The essential reason is that the RSSI information is available on all participants. 

Further, from the above discussion, we can see that the number of equations is much larger than the number of unknowns. As we want to utilize all measurements for a balanced solution to reduce error via variation across all network nodes, we design the optimization objective function to minimize the total deviation from all equations, which is reviewed below.

\begin{equation}
\label{eq:eq11}
\widehat{\gamma}_i,\widehat{d}_{12}=
\argmin_{\gamma_i,d_{12}}\sum\limits_{i,j}^{}(r_{ij}-(R_i-10\gamma_ilog(p_{ij}d_{12})))^2
\end{equation}

%With each pair of equations for node $i$, we can obtain one result of 
Path loss exponent $\gamma_i$ is solved for each node to accommodate different surrounding environments. After $d_{12}$ is obtained, all the distances between each pair of nodes can be derived with topology description.

%As we have the geographical description, we could avoid the rotation Here, we assume each network node know its own $P_i$ and path loss exponent $\gamma$ is set to 3 for simplicity. with more devices > 5, we could resolve different path loss exponent $\gamma_i$  for each node $i$
% what's the necessity/difference from other systems/ why it is good for our target application?
% the design two components of the solution. 
%1. the self-localizing requires minimum hardware - 2 nodes, no infrastructure.
%2. the error of first algorithm mainly comes from the initial pair localization error larger than 2m. 
%3. the second algorithm mainly utilize the high accuracy of aoa, and estimating all distance in a single parameter - scale size. 
%the insight is that the first algorithm didn't utilize all rssi information. 
% 
% \sysname\ could also cooperate with existing network infrastructure to provide better accuracy of localization. Think about (scenario)...